# A computational Simulation of Steady Natural Convection in an H-Form Cavity

Mohamed Loukili[1], Kamila Kotrasova[2], Denys Dutykh[3]

[1] Faculty Ben M'sik of Sciences, Hassan II University, Morocco
[2] Faculty of Civil Engineering, Technical University of Kosice, Slovak Republic
[3] Univ. Grenoble Alpes, Univ. Savoie Mont Blanc, CNRS, LAMA, 73000 Chambéry, France

md.loukili@gmail.com

**Abstract.** The simulation of natural convection problem based on the Galerkin finite-element method, with the penalty finite-element formulation of the momentum balance equation, is exploited for accurate solutions of equations describing the problem of H-Form cavity differentially heated side walls. The cavity is occupied by the air whose Prandtl number is $Pr$=0.71, the fluid is assumed to be steady, viscous and incompressible within thermal convection. A numerical investigation has been made for Rayleigh numbers ranging from 10 to $10^6$ for three cases of total internal height aspects of H-Form cavity: 0%, 50%, and 85%. Firstly, the goal is to validate the numerical code used to resolve the equations governing the problem of this work. For that, we present a comparison between the profiles at the point (0.5, 0) for the $u$-component, and $u$-component obtained in previous work for simple square cavity. Further, a comparison of the averaged Nusselt number with previous works for simple square cavity is realized in order to ensure the numerical accuracy, and the validity of our considered numerical tool. Secondly, the objective is to investigate on the hydrodynamic effects of Rayleigh number for different total internal height aspects of H-Form cavity on the dynamics of natural convection. Shortly after, the ambition is to assess the heat transfer rate for different Rayleigh number for three cases of internal height aspects.

## 1 Introduction

The natural convection flows have been the subject of many investigations due to its basic importance in numerous industrial and natural processes, such as: solar collectors, cooling of electronic equipment, energy storage systems, air conditioned system in buildings, thermal insulation, and fire propensity control in buildings [1–3]. Further, the natural convection mechanism within square cavity has been attracting the interest of many researchers for several decades [4-9], because of its importance and the great number of applications in various fields and industries. In 1983, De Vahl Davis and G. Jones [7] corroborated the accuracy of the benchmark solution and offered a basis solution to ensure the validity of new contribution in natural convection problem. Next, two sequences of simulations were studied to deal with the impact of buoyancy force on mass transfer rate, and to get the effect of Lewis number on fluid motion [10]. After that, the transient features of natural convection flow in the partitioned cavity were discussed [9], and the authors shed light on three different stages (initial, transitional, and steady) as the time scale was considered. Then, a membrane was introduced to the square cavity to study different aspect of natural convection for different Rayleigh numbers [8]. Two years before, the mechanism of natural convection of air within a square cavity with inferior walls and active side was experimentally and computationally investigated [11]. This last paper showed that the average and maximum velocity, with small values characteristics of natural convection, rise with Rayleigh and even with the angle and attain an extreme value for Ra maximum. More recently, A. Mazgar et al. [12] addressed the impact of gas radiation on laminar natural convection flow inside a square cavity that has an internal heat source, and emphasized the entropy generation. They concluded a significant finding from this investigation is that radiative influence has a key role in the acceleration of the vortices and affording either a homogenizing impact on temperature fields.

Earlier investigators have theoretically and experimentally addressed many aspects of convective heat transfer in cavity enclosures involving conjugate heat transfer effects [13], nanofluids and entropy generation [14], magnetic field effects [15], cavity filled with porous media [16] and the presence of a solid partition [17]. Furthermore, many researchers have studied different geometry aspects of convective heat transfer in simple enclosures, such as the geometry of triangular shape [18], C-shape [19], concentric annulus [20], hemispherical shape [21–23], and parallelo-grammic shape [24].

In this work, the objective is to study the natural convection inside H-Form cavity, and to assess the impact of Rayleigh number for three cases of internal height aspects of H-Form cavity: 0%, 50%, and 85% on the fluid flows, and the heat transfers. To realize this work, the paper is illustrated in five sections: after the introduction in the first section, the problem statement and solutions procedures are clearly described in the second section. Shortly after, we will interest to corroborate the accuracy and validate the numerical code used in this work. Next, we shed the light on the hydro-dynamic impact of Rayleigh number, and the size of H-Form cavity on; flow pattern and heat transfer mechanism inside the defined computational domain. Finally, we finish our manuscript by outlining the main conclusions and perspectives of this study.

## 2    Problem statement and solutions procedures

The studied computational domain is a two-dimensional H-Form cavity, the total internal height aspects of H-Form cavity are stands for *2l* (see Fig.1). The cavity is occupied by the air whose Prandtl number is *Pr*=0.71, the fluid considered is assumed to be steady, viscous and incompressible within thermal convection. The upstream and downstream sides are differentially heated with fixed temperatures, whereas the other sides are considered adiabatic and the velocities on every side are zero. The studied configuration is sketched in Fig.1.

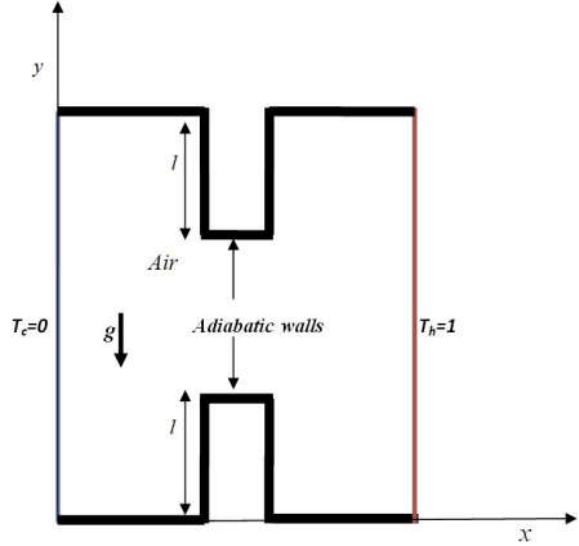

**Fig.1**. Physical domain

The non-dimensional equations describing the steady natural convection flow, are expressed as

$$\frac{\partial u}{\partial x} + \frac{\partial v}{\partial y} = 0,$$ (1)

$$u\frac{\partial u}{\partial x} + v\frac{\partial u}{\partial y} = -\frac{\partial P}{\partial x} + Pr\left(\frac{\partial^2 u}{\partial x^2} + \frac{\partial^2 u}{\partial y^2}\right),$$ (2)

$$u\frac{\partial v}{\partial x} + v\frac{\partial v}{\partial y} = -\frac{\partial P}{\partial y} + Pr\left(\frac{\partial^2 v}{\partial x^2} + \frac{\partial^2 v}{\partial y^2}\right) + Ra\,Pr\,T,$$ (3)

$$u\frac{\partial T}{\partial x} + v\frac{\partial T}{\partial y} = \frac{\partial^2 T}{\partial x^2} + \frac{\partial^2 T}{\partial y^2}.$$ (4)

and the boundary conditions are expressed as

$$\begin{cases} u = v = 0 \\ T_c = 0 \end{cases} \qquad \text{for } x=0, \text{ and } y \in [0\ 1]$$ (6)

$$\begin{cases} u = v = 0 \\ T_h = 1 \end{cases} \qquad \text{for } x=1, \text{ and } y \in [0\ 1]$$ (7)

$$\begin{cases} u = v = 0 \\ \frac{\partial T}{\partial y} = 0 \end{cases} \qquad \text{for } (y=0 \text{ and } y=1), \text{ and } x \in [0\ 1]$$ (8)

the different parameters are given as follows: $u$ and $v$ denote the non-dimensional velocities, $T$ is the non-dimensional temperature, $Ra$ stands for Rayleigh number, $Pr$ stands for Prandtl number.

The momentum and energy balance equations [Eqs. (2)–(4)] are solved using Galerkin finite element method. The continuity equation [Eq. (1)] will be used as a constraint due to mass conservation and this constraint may be used to obtain the pressure distribution [25-27]. In order to solve Eqs. (2)–(3), we use the penalty finite element method where the pressure $P$ is eliminated by a penalty parameter $\delta$ and the incompressibility criteria given by Eq. (1) results in

$$p = -\delta\left(\frac{\partial u}{\partial x} + \frac{\partial v}{\partial y}\right), \tag{9}$$

the continuity equation (1) is satisfied for numerous values of $\delta$, the typical value of $\delta$ which affords consistent solution is $\delta = 10^7$ [27]. Then, the governing equations (2) and (3) are reduced to

$$u\frac{\partial u}{\partial x} + v\frac{\partial u}{\partial y} = \delta\frac{\partial}{\partial x}\left(\frac{\partial u}{\partial x} + \frac{\partial v}{\partial y}\right) + Pr\left(\frac{\partial^2 u}{\partial x^2} + \frac{\partial^2 u}{\partial y^2}\right), \tag{10}$$

$$u\frac{\partial v}{\partial x} + v\frac{\partial v}{\partial y} = \delta\frac{\partial}{\partial y}\left(\frac{\partial u}{\partial x} + \frac{\partial v}{\partial y}\right) + Pr\left(\frac{\partial^2 v}{\partial x^2} + \frac{\partial^2 v}{\partial y^2}\right) + Ra\,Pr\,T. \tag{11}$$

The penalty finite-element method is a standard technique for solving incompressible viscous flows, the Galerkin finite element method yields the nonlinear residual equations for Eqs. (10)–(11). By adopting biquadratic basis functions with three points Gaussian quadrature in order to evaluate the integrals in the residual equations, the numerical procedure details are depicted in the reference [27].

## 3    Computational validation

Firstly, the goal is to ensure the validity and verify the accuracy of the numerical technique adopted to resolve the problem of this work. In Fig. 2, we present a comparison between the profiles at the point (0.5, 0) for the $u$-component for simple square cavity (a), and $u$-component obtained in the reference [27] at the mid-width of the cavity (b) for the same boundary conditions. The results of the Fig. 2 are shown a good agreement with the results obtained by Mousa [27] for Rayleigh number $Ra$ ranging from 10 to $10^6$.

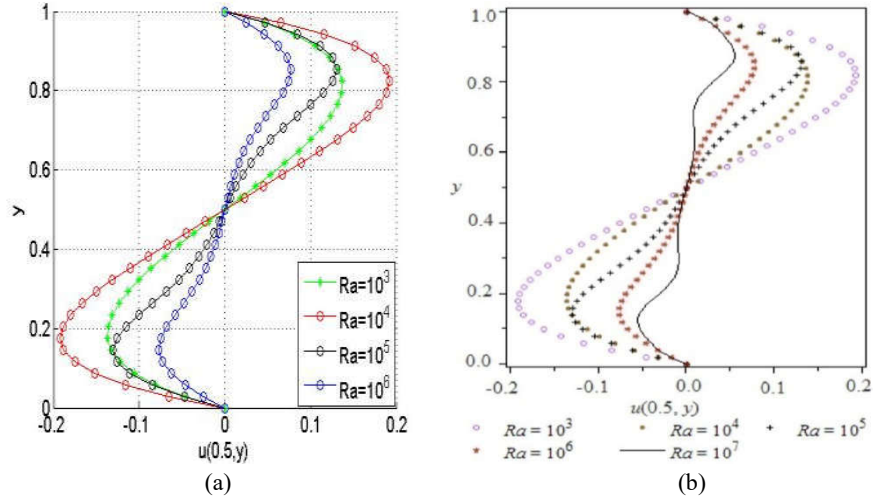

**Fig.2.** (a) Distribution of *u*-velocity component at mid-width of simple square cavity, (b) distribution of *u*-velocity component at mid-width obtained in [27], for various *Ra* values.

Secondly, to ensure numerical validity of our numerical code, we illustrate the comparison of the averaged Nusselt number with previous works for square cavity for different values of Rayleigh number, see the Table 1. Thereafter, we calculate the Nusselt's deviation $D_v$ of the present work from the following references (Fusegi et al [28], De vahl Davis [7], and Barakos et al [29]).

The Nusselt's deviation $D_v$ is defined as

$$D_v = \frac{|Nu^P - Nu^R|}{Nu^R}$$

$Nu^P$ stands for the Nusselt number of the present work,
$Nu^R$ stands for the Nusselt number of the designed reference.

**Table 1.** The Nusselt's deviation concerning square cavity for different Rayleigh numbers.

| $Ra$ | Present work | Fusegi et al [28] | | | De vahl Davis [7] | | Barakos et al [29] | |
|---|---|---|---|---|---|---|---|---|
| | $Nu^p$ | $Nu^R$ | $Dv(\%)$ | | $Nu^R$ | $Dv(\%)$ | $Nu^R$ | $Dv(\%)$ |
| $10^3$ | 1.117 | 1.105 | 1,09 | | 1.118 | 0,09 | 1.114 | 0,27 |
| $10^4$ | 2.244 | 2.302 | 2,52 | | 2.243 | 0,04 | 2.245 | 0,04 |
| $10^5$ | 4.520 | 4.646 | 2,71 | | 4.519 | 0,02 | 4.510 | 0,22 |
| $10^6$ | 8.820 | 9.01 | 2,11 | | 8.799 | 0,24 | 8.806 | 0,16 |

The Table 1 addresses the comparison of the averaged Nusselt along the hot wall, the results presented show an excellent agreement between the present results and the benchmark solutions [7, 28, and 29] for all values of Rayleigh number.

## 4    Results and discussion

The present section puts the light on the hydrodynamic impact of Rayleigh number, for three cases of total internal height aspects of H-Form cavity on both; flow pattern and heat transfer mechanism inside the defined computational domain. Firstly, The Fig. 2 highlights streamlines for Rayleigh numbers ranging from 10 to $10^6$ for three cases of total internal height aspects: 0%, 50%, and 85%.

*Ra*=10

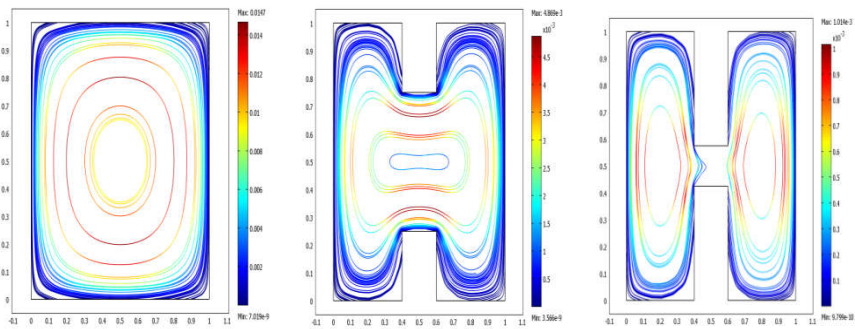

*Ra*=$10^2$

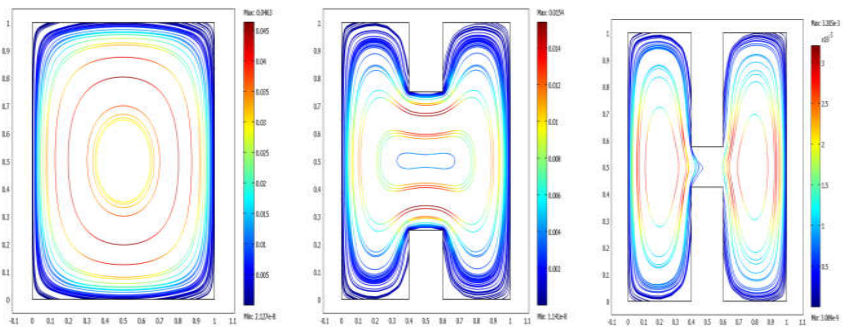

*Ra*=10³

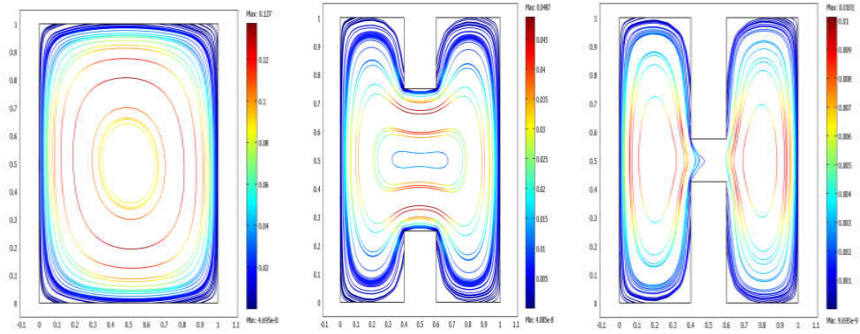

*Ra*=10⁴

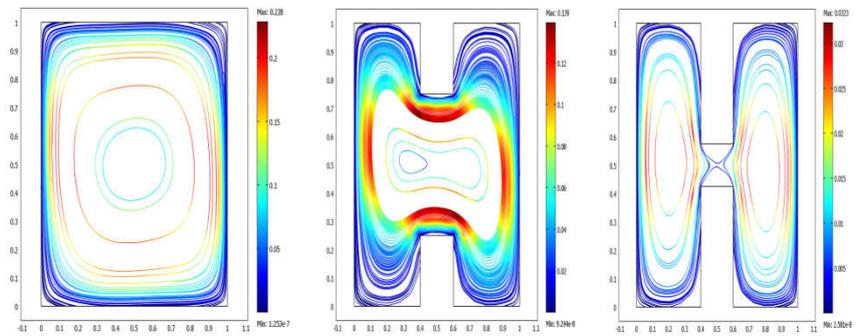

*Ra*=10⁵

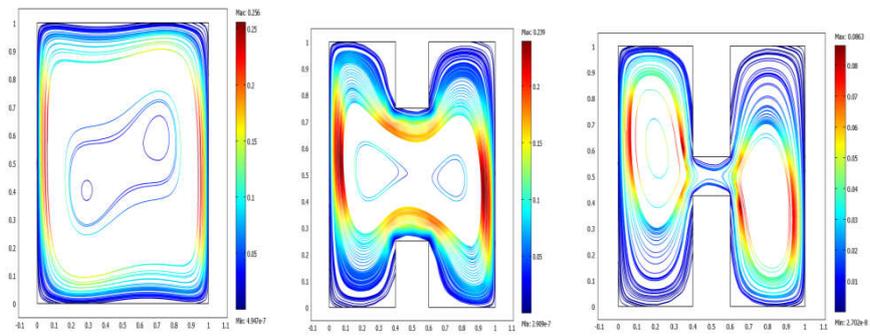

$Ra = 10^6$

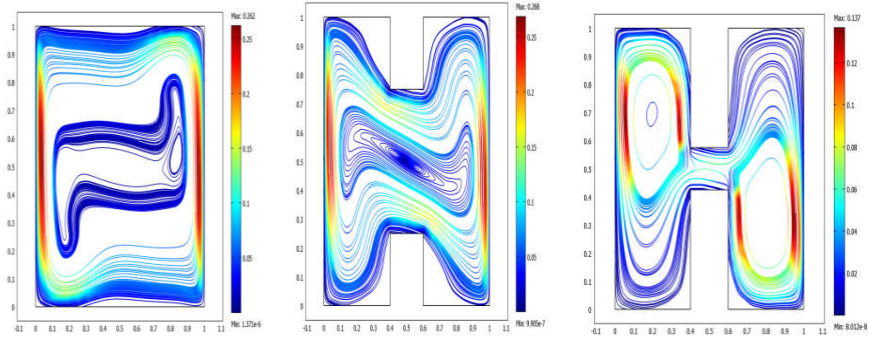

**Fig.3**. Streamlines in H form cavity for different total internal height aspects and for Rayleigh number ranging from 10 to $10^6$.

Based on the Fig. 3 the outcomes reveal that the Rayleigh number is the major parameter influencing the flow in the case of simple square cavity. In detail, concerning the low values of Rayleigh number ($Ra$ from 10 to $10^3$ ) a vortex is created in the center, when Rayleigh number growths to $Ra = 10^4$ the vortex seems to be elliptic, as Rayleigh number increases to $Ra = 10^5$ two vortices are generated at the center giving space to the third vortex to develop at $Ra = 10^6$.

Thereafter, for the cavity with 50% of total internal height aspects we remark in general that the velocities are less important than the flow in simple square cavity, which is totally logical regarding the dynamics of the flows-structure interactions. The figures show that Rayleigh number ranging from 10 to $10^3$ the streamlines appear elliptic, for $Ra = 10^4$ the velocities are higher at the center creating a vortex and giving space to the second vortex to develop at $Ra = 10^5$, the two vortices meet at center creating a large vortex. By increase of the buoyant force via increase in the Rayleigh number, the flow intensity increases and the streamlines closes to the side walls

Next, for the cavity with 85% of total internal height aspects we remark that the circulations of fluid flow are blocked in each block of H-form geometry creating vortices at each block separately. Consequently, we notice that the Rayleigh number hasn't a strong effect on the characteristics of the fluid flow, and Rayleigh number effect starts to appear till $Ra = 10^4$ where the two vortices of each block are start to be in contact, once Rayleigh number achieve $Ra = 10^5$ the exchange between two columns becomes important. Lastly, for $Ra = 10^6$ vortices are more vital, improving stratifications at the left top and downright of each block of H-form cavity.

Secondly, the Fig. 4 presents the isotherms for Rayleigh numbers ranging from 10 to $10^6$ for three cases of total internal height aspects: 0%, 50%, and 85%. We remark, in the case of simple square cavity 0% of total internal height aspects, the heat transfer mechanism changes as a function of Rayleigh number. To clear up, we notice that from $Ra = 10$ to $Ra = 10^3$ the isotherms appear vertical, when Rayleigh num-ber increases to $Ra = 10^4$ the heat transfer mechanism changes from conduction to convection, as Rayleigh number raises the isotherms are no longer vertical only inside the very thin boundary layers. Whereas, in the case of cavity with 50% and 85% of total internal height aspects we observe that the circulation of the fluid flow becomes slower compared to the case of simple square cavity, which explain the slow action of the heat transfer mechanism when the total internal height aspects increase.

Next, for the cavity with 85% of total internal height aspects we remark that the Rayleigh number is not dominant till it reaches $10^5$ and then the heat transfer mechanism starts to change from conduction to convection, at $Ra = 10^6$ the process of convection is no longer fast due to the blockage of the circulation of the fluid flow inside each block of H-form cavity.

*Ra*=10

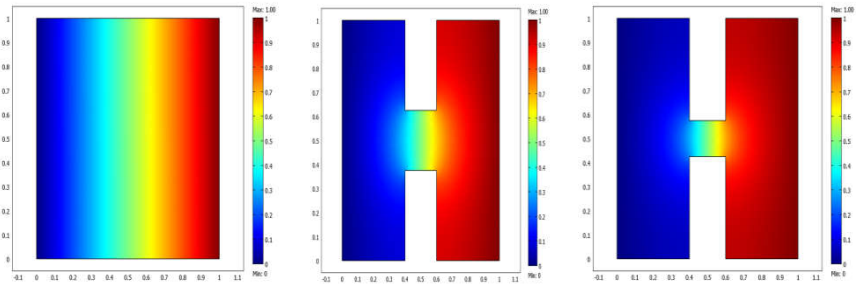

*Ra*=$10^2$

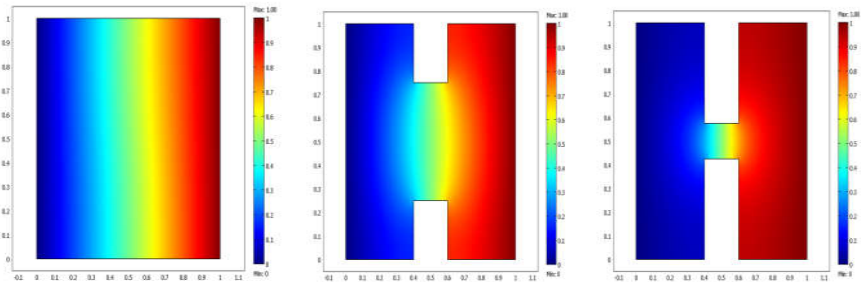

*Ra*=$10^3$

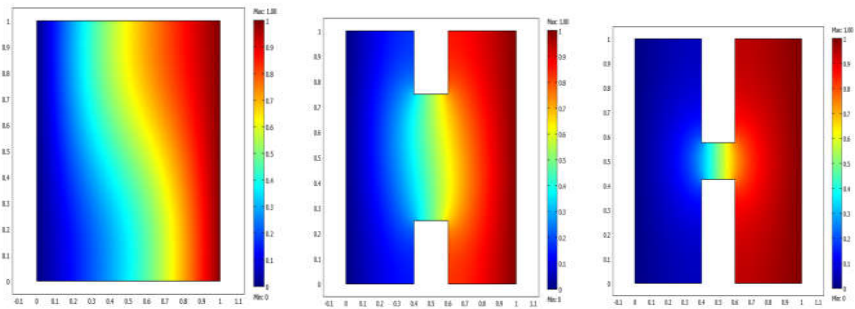

$Ra=10^4$

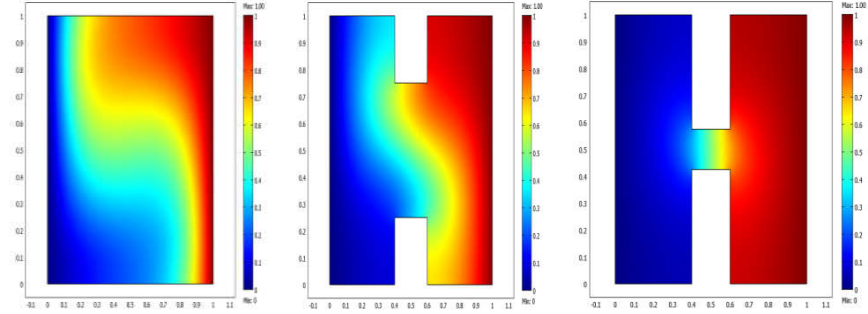

$Ra=10^5$

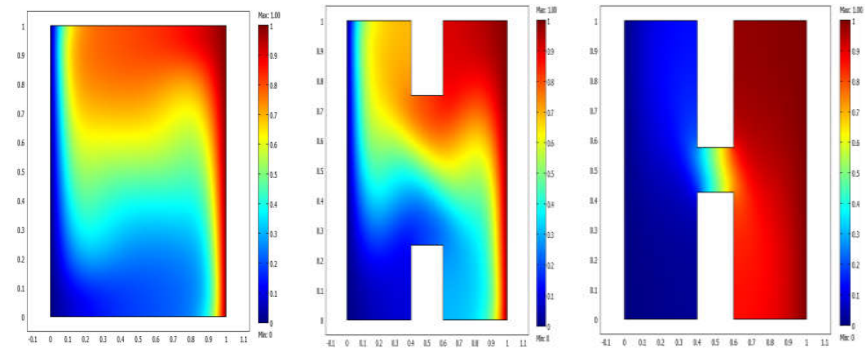

$Ra=10^6$

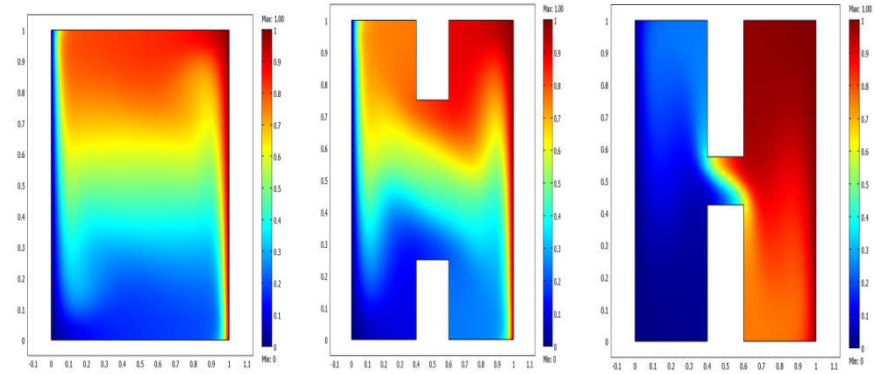

**Fig.4**. Isotherms in H form cavity for different total internal height aspects and for Rayleigh number ranging from 10 to $10^6$.

In this subsection, the local Nusselt number along the hot wall of the cavity is presented in Fig. 5 for a wide range of Rayleigh numbers for three cases of total internal height aspects. The goal is to analyze the effects of the increase of total internal height aspects on the local Nusselt number along the hot wall of the cavity at various Rayleigh numbers.

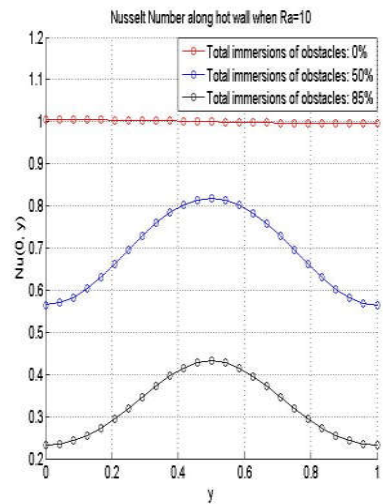

(a)

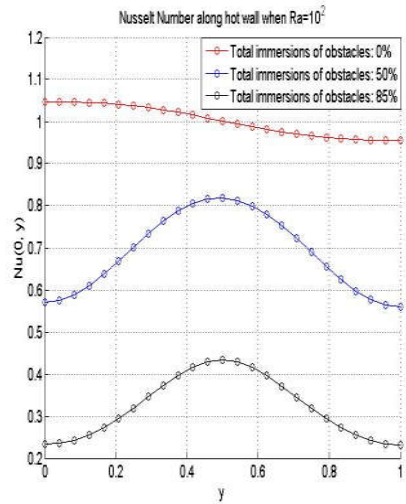

(b)

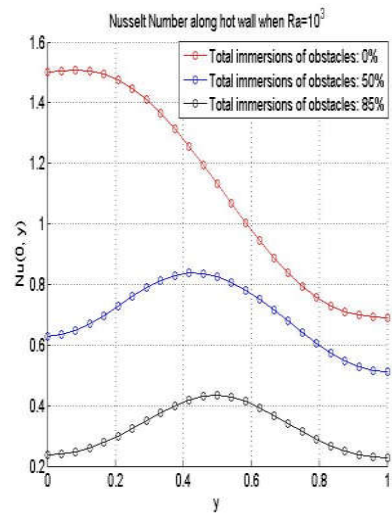

(c)

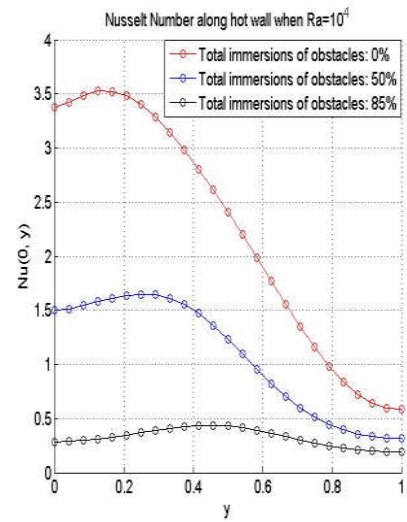

(d)

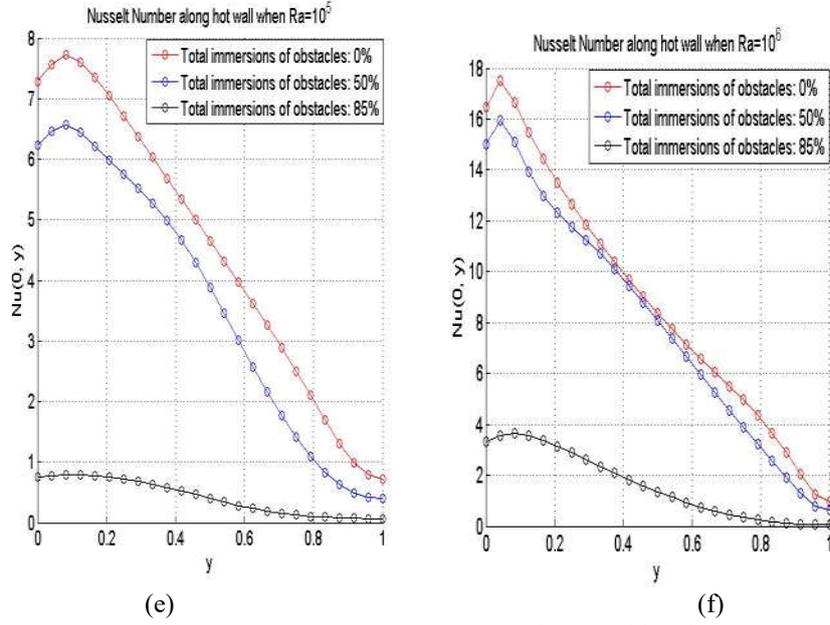

(e)                                                    (f)

**Fig.5**. Variation of the local Nusselt number *Nu* with total internal height aspects at various Rayleigh numbers. **a** $Ra = 10$, **b** $Ra = 10^2$, **c** $Ra = 10^3$ **d** $Ra = 10^4$, **e** $Ra = 10^5$, **f** $Ra = 10^6$.

The outcomes of the Fig. 5 present the local Nusselt number along the hot wall of the cavity at various Rayleigh numbers for different total internal height aspects. For all aspect ratios and for different Rayleigh numbers the maximum local Nusselt number occurs at the lower end of the hot wall i.e. $y = 0$. Further, the local Nusselt number decreases as total internal height aspects increases, and then the heat transfer rate decreases with the increase of the total internal height aspects, this is due to the fact that when the total internal height aspects increases the fluid is damped and hence rate of free convection decreases. Furthermore, when the Rayleigh numbers increase the cold fluid moves to hot wall and hence maximum temperature gradient occurs at this region, the cold fluid ascends adjacent to the hot wall, then the fluid temperature increases subsequently the local Nusselt number decreases. For the case of simple square cavity, the local Nusselt profile departs from his vertical position and then the process of the conviction is started at $Ra = 10^4$. Moreover, for the cases of total internal height aspects: 50% and 85% and when Rayleigh numbers ranging from $Ra = 10$ to $Ra = 10^3$ the location at which the maximum local Nusselt number occurs is about $y = 0.5$. In addition, when Rayleigh numbers ranging from $Ra = 10^5$ to $Ra = 10^6$ for all aspect ratios, maximum rate of heat transfer occurs at about $y < 0.2$ of the hot wall, this position approaches the lower part of the hot wall as $Ra$ increases.

## 5 Conclusion and perspectives

In this work, the natural convection in H-form cavity has been studied numerically for different parameters influencing the flow and heat transfer. In details, we have assess the impact of Rayleigh number for three cases of total internal height pects: 0%, 50%, and 85%. From the foregoing discussion, we remark in the case of simple square cavity (0% of total internal height aspects) that Rayleigh number is the dominant parameter on the heat transfer mechanism and fluid flows. In the case of H-Form cavity with 50% and 85% total internal height aspects, and Rayleigh numbers ranging from $Ra = 10$ to $Ra = 10^3$ the maximum of rate transfers are occurs at the centre. Rayleigh number does not influencing the heat transfer till it reaches $10^4$ for 50% of total internal height aspects and $10^5$ for 85%. In general, for all aspect ratios, the heat transfer rate increases with Rayleigh number and decreases with the increase of the total internal height aspects. Furthermore, as the total internal height aspects increases the circulation of the fluid flow becomes slower, which causes to decreases effect of heat transfers rate. As perspectives, we endeavor to study the natural convection flows in H-form cavity using meshless methods [30-33], and spectral methods [34] that have been proven a strong efficiency in many nonlinear and engineering fields [30-33].